\documentclass[epj]{svjour}
\usepackage{graphicx}
\usepackage{psfrag,graphics}
\usepackage{amsmath}
\usepackage{subfigure}
\usepackage{lineno}
\usepackage{mathtools}
\usepackage{setspace}
\usepackage{url}
\begin{document}
\title{Inclusive dielectron spectra in p+p collisions at 3.5 GeV}
\author{G.~Agakishiev$^{6}$, A.~Balanda$^{3}$, D.~Belver$^{17}$, A.~Belyaev$^{6}$, 
A.~Blanco$^{2}$, M.~B\"{o}hmer$^{9}$, J.~L.~Boyard$^{15}$, P.~Cabanelas$^{17}$, E.~Castro$^{17}$, 
J.C.~Chen$^{8}$
% ...
} 


\institute{
(HADES collaboration) \\\mbox{$^{1}$Istituto Nazionale di Fisica Nucleare - Laboratori Nazionali del Sud, 95125~Catania, Italy}\\
\mbox{$^{2}$LIP-Laborat\'{o}rio de Instrumenta\c{c}\~{a}o e F\'{\i}sica Experimental de Part\'{\i}culas , 3004-516~Coimbra, Portugal}\\
\mbox{$^{3}$Smoluchowski Institute of Physics, Jagiellonian University of Cracow, 30-059~Krak\'{o}w, Poland}\\
\mbox{$^{4}$GSI Helmholtzzentrum f\"{u}r Schwerionenforschung GmbH, 64291~Darmstadt, Germany}\\
\mbox{$^{5}$Institut f\"{u}r Strahlenphysik, Helmholtz-Zentrum Dresden-Rossendorf, 01314~Dresden, Germany}\\
\mbox{$^{6}$Joint Institute of Nuclear Research, 141980~Dubna, Russia}\\
\mbox{$^{7}$Institut f\"{u}r Kernphysik, Goethe-Universit\"{a}t, 60438 ~Frankfurt, Germany}\\
\mbox{$^{8}$Excellence Cluster 'Origin and Structure of the Universe' , 85748~Garching, Germany}\\
\mbox{$^{9}$Physik Department E12, Technische Universit\"{a}t M\"{u}nchen, 85748~Garching, Germany}\\
\mbox{$^{10}$II.Physikalisches Institut, Justus Liebig Universit\"{a}t Giessen, 35392~Giessen, Germany}\\
\mbox{$^{11}$Istituto Nazionale di Fisica Nucleare, Sezione di Milano, 20133~Milano, Italy}\\
\mbox{$^{12}$Institute for Nuclear Research, Russian Academy of Science, 117312~Moscow, Russia}\\
\mbox{$^{13}$Frederick University, 1036~Nicosia, Cyprus}\\
\mbox{$^{14}$Department of Physics, University of Cyprus, 1678~Nicosia, Cyprus}\\
\mbox{$^{15}$Institut de Physique Nucl\'{e}aire (UMR 8608), CNRS/IN2P3 - Universit\'{e} Paris Sud, F-91406~Orsay Cedex, France}\\
\mbox{$^{16}$Nuclear Physics Institute, Academy of Sciences of Czech Republic, 25068~Rez, Czech Republic}\\
\mbox{$^{17}$Departamento de F\'{\i}sica de Part\'{\i}culas, Univ. de Santiago de Compostela, 15706~Santiago de Compostela, Spain}\\ 
\\
\mbox{$^{A}$ also at Lawrence Berkeley National Laboratory, ~Berkeley, USA}\\
\mbox{$^{B}$ also at ISEC Coimbra, ~Coimbra, Portugal}\\
\mbox{$^{C}$ also at ExtreMe Matter Institute EMMI, 64291~Darmstadt, Germany}\\
\mbox{$^{D}$ also at Technische Universit\"{a}t Darmstadt, ~Darmstadt, Germany}\\
\mbox{$^{E}$ also at Technische Universit\"{a}t Dresden, 01062~Dresden, Germany}\\
\mbox{$^{F}$ also at Dipartimento di Fisica, Universit\`{a} di Milano, 20133~Milano, Italy}\\
\mbox{$^{G}$ also at Dipartimento di Fisica Generale and INFN, Universit\`{a} di Torino, 10125~Torino, Italy}\\
\\
\mbox{$^{*}$  \emph{Corresponding author:} Rustamov@Physik.uni-frankfurt.de}\\
}

%
%
%

\date{Received: 15.12.2011 / Revised version: date}

%
\abstract{
We present the inclusive invariant-mass, transverse momentum and rapidity distributions of dielectrons (e$^{+}$e$^{-}$ pairs) in p+p interactions at 3.5 GeV beam kinetic energy. In the vector-meson mass region, a distinct peak corresponding to direct $\omega$ decays is reconstructed with 2$ \%$ mass resolution. The data is compared to predictions from three model calculations. Due to the large acceptance of the HADES apparatus for $e^{+}e^{-}$ invariant masses above 0.2 GeV/$c^{2}$ and for transverse pair momenta p$_{t}$ $ <$ 1 GeV/$c$, acceptance corrections are to a large extent  model independent. This allows us to extract from dielectron data for the first time at this energy the inclusive production cross sections for light vector mesons. Inclusive production cross sections for $\pi^o$ and $\eta$ mesons are also reported. The obtained results will serve as an important reference for the study of vector meson production in proton-nucleus and heavy-ion collisions. Furthermore,  using this data, an improved value for the upper bound of the branching ratio for direct $\eta$ decays into the electron-positron channel is obtained. 
\PACS{
      {25.40.Ep, 13.40.Hq, 13.60.Le, 13.60.Rj}{}
     } 
} 

\authorrunning{HADES collaboration}
\titlerunning{Inclusive dielectron spectra in p+p collisions at 3.5 GeV}

\maketitle

\section{Introduction}
\label{sec:Introduction}

The High Acceptance Di-Electron Spectrometer (HADES) \cite{Aga:2009} is operated at the GSI Helmholtzzentrum 
f\"{u}r Schwerionenforschung in Darmstadt,
Germany. One of the main physics goals of HADES is to investigate spectral modifications of  light vector mesons in strongly interacting matter. The question is how the low-energy QCD spectrum, which is experimentally  known in the vacuum in terms
of hadron spectra, will change when this vacuum is heated and filled up with color charges. Spectral modifications of hadrons (encoded, e.g., by 
changes of their masses and decay widths) in hot and/or dense matter are often discussed in the context of the restoration of the broken chiral symmetry. 
Detailed investigations, however, reveal that the link between hadron properties and  QCD symmetries is not as direct as
originally envisaged \cite{Metaq:Leupold,Rapp:Wambach}. 

In order to search for in-medium effects, results on electron-positron invariant mass spectra from proton induced reactions on nuclei and from heavy-ion collisions  should be compared systematically and complemented with insights gained from photo-induced reactions. The interpretation of nuclear data requires firm knowledge of the corresponding data from proton-proton (p+p) collisions. These investigations are also important  for transport model calculations, as the dilepton spectra from elementary interactions serve as input to these codes.  Given this motivation, the HADES collaboration has set up an experimental program to measure dilepton spectra in elementary collisions. First results were discussed in \cite{Aga:2010}. Here we present the inclusive dielectron spectra measured in p+p collisions at 3.5 GeV kinetic beam energy.

At this energy the dominant mechanisms for hadron production are still controversially discussed and inclusive production cross sections are not yet measured. In fact, various production scenarios  are assumed in different transport codes: While in HSD\footnote{Hadron-String Dynamics}~\cite{HSD:web} and GiBUU\footnote{Giessen Boltzmann-Uehling-Uhlenbeck project}~\cite{GiBUU:web} hadrons are produced at this beam energy through string fragmentation  \cite{Lund:1983}, in UrQMD\footnote{Ultrarelativistic Quantum Molecular Dynamics}~\cite{UrQMD:web} the decays of nucleon resonances are the sources of final state particles, in particular of mesons.

Another uncertainty concerns the Dalitz decays of baryon resonances (R $\to$ N e$^{+}e^{-}$). Here two different aspects are a matter of discussion. 
First, the R Dalitz decay process, i.e. its decay into a nucleon and a massive photon with the subsequent decay of the latter into a dielectron depends on the electromagnetic structure of the N-R transition vertex. In case of the $\Delta(1232)$ resonance this transition vertex depends on three independent helicity amplitudes corresponding to three helicity states of the massive photon and two of the nucleon. Equivalently, one can describe this vertex by three independent transition form factors built up from the helicity amplitudes. In this decay process the squared four-momentum of the virtual photon equals the squared invariant mass of the lepton pair and therefore is a positive quantity.  Such a photon transfers energy and  is  said to be a time-like photon\footnote{We use the following convention of the squared 4-momenta:  $p^{2} = p^{\mu}(E,\vec{p})p_{\mu}(E,-\vec{p}) = E^{2}-|\vec{p}|^{2}$, i.e.,  positive $p^{2}$ is called time-like, and negative $p^{2}$ is correspondingly space-like.}. There are also other processes (e.g. pion electro-production) where the virtual photon transfers momentum, but not energy, and is then referred to as space-like. 
While in case of space-like photons the above mentioned transition form factors have been measured in quite a wide range of four-momentum transfer, for time-like photons their mass dependence is not settled yet. The mass dependence of the form factors is usually modeled within the Vector Meson Dominance (VMD) model. In this picture the virtual photon couples to the nucleon through intermediate vector meson ($\rho$, $\omega$, $\phi$) states. However, it is known that the standard VMD model of Sakurai \cite{Sakurai:1960,Sakurai:1969} significantly overestimates the radiative (R $\to$ N$\gamma$) branching ratios once the coupling constant is extracted from the corresponding mesonic decays. The modified VMD model of Kroll et al. \cite{Kroll:1967}  allows to fix the $\rho$N and $\gamma$N coupling constants independently. The asymptotic behavior of the transition form factors predicted by both models disagrees however with the outcome of quark counting rules \cite{eVMD:2002}. 
On the other hand, the authors of \cite{eVMD:2002}  demonstrated in their framework of extended VMD that the incorporation of higher vector meson states  resolves the problem between photon and $\rho$ meson branchings of the nucleon resonances. Yet another VMD model variant for the nucleon form factors was proposed by Iachello \cite{Iacello:1973}. It describes simultaneously  the nucleon space-like and time-like form factors as well as space-like N-$\Delta$ transition form factors \cite{Iacello:2004,Iacello:2005}.

A second aspect of resonance decays to consider is the parametrization of the mass dependent resonance width. Various prescriptions are used in model  calculations (for details see section 3 and \cite{Kemp:2009}) which differ at high resonance masses. As a consequence, the resulting dielectron yield from the resonance Dalitz decays has a large uncertainty, as shown for the $\Delta(1232)$ in \cite{Kemp:2009}.

Our precision data offer hence a unique possibility to address the above-mentioned problems.

This paper is organized as follows. After giving detailed information about the collected data and analysis chain in section~\ref{sec:Data}, we report in section~\ref{sec:Models} on a comparison of the data with model calculations.
The sensitivity of the data to the N-$\Delta$ electromagnetic transition vertex is also discussed in this section. The extraction of $\pi^{o}$, $\eta$, $\Delta$(1232), $\rho$ and $\omega$ cross sections is discussed in section~\ref{sec:Cross1}. 
The data allow to improve the upper bound of the direct $\eta \to e^{+}e^{-}$ decay, as discussed in section~\ref{sec:etaDir}.

\section{The data}
\label{sec:Data}

In the experiment (see \cite{Aga:2009} for a detailed description of HADES), a proton beam of $10^{7}$ particles/s with a kinetic energy of 3.5 GeV was incident on a 4.4 cm long liquid hydrogen target \cite{Rust:2010}. The data readout was started upon a first-level trigger (LVL1) decision. Depending on the reaction channel of interest, two different settings of the LVL1 trigger were required: (i) a charged-particle multiplicity MULT $\geq$ 3 to enhance inclusive dielectron production and (ii) MULT $\geq$ 2 with hits in opposite sectors ($\Delta \phi = 180^{o} \pm 60^{o}$) of the time-of-flight detectors to enrich elastic p+p events used for the absolute normalization of the dielectron data. The LVL1 was followed by a second-level trigger (LVL2) requesting at least one electron candidate recognized in the Ring-Imaging Cherenkov Detector (RICH) and time-of-flight/pre-shower detectors \cite{Aga:2009}. All events with positive LVL2 trigger decision and every third LVL1 event, irrespective of the LVL2 decision, were recorded, yielding a total of 1.17$\times10^{9}$ events.

\begin{figure}[htb]
 {\includegraphics[width=0.99\linewidth,clip=true]
      {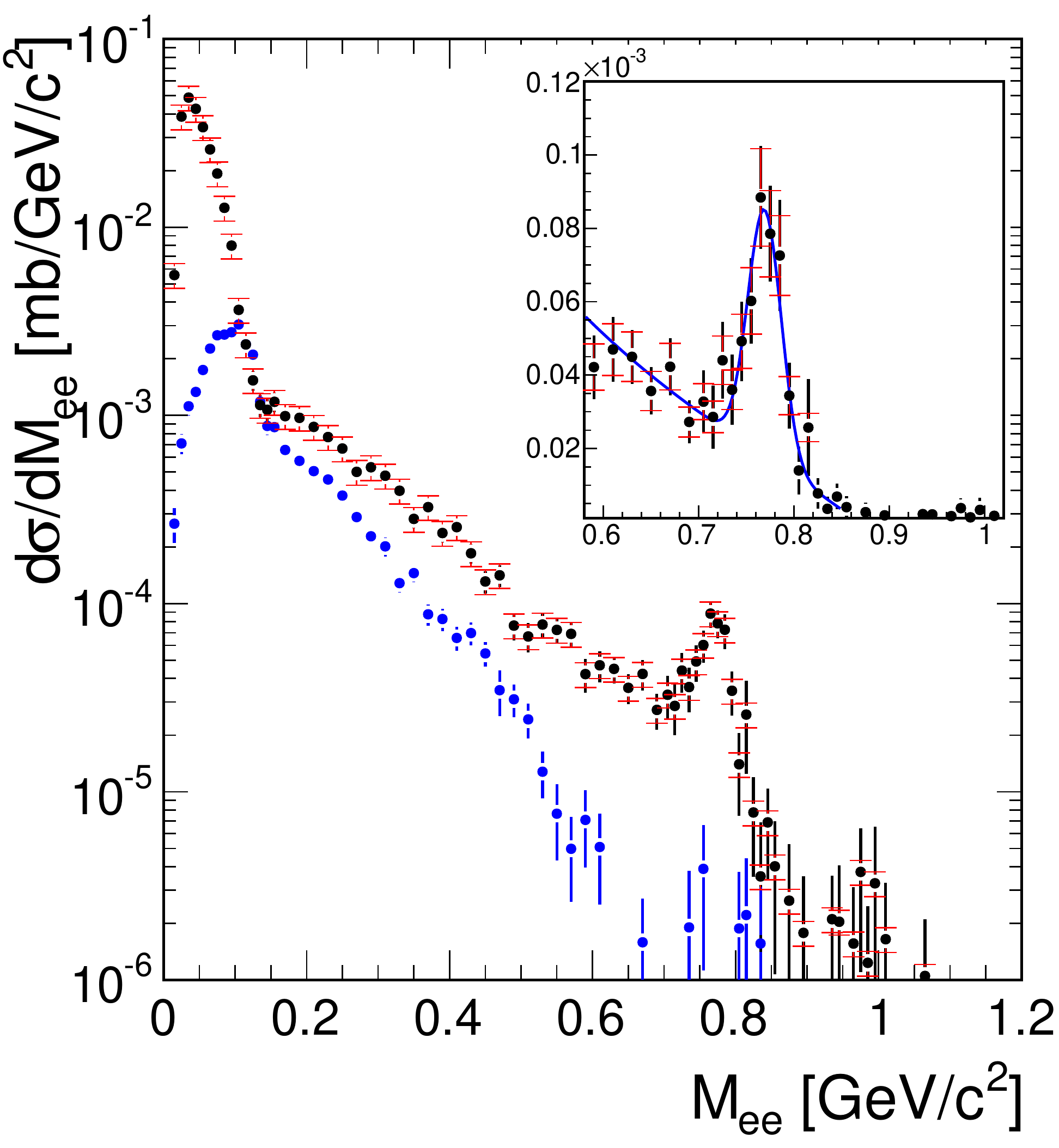}
}
\caption{(color online) Efficiency corrected inclusive invariant-mass distribution of dielectrons inside the geometrical acceptance of HADES of signal (black dots) e$^{+}$e$^{-}$ pairs after combinatorial background (blue dots) subtraction for p(3.5GeV)+p collisions. In the right upper part of the figure the $\omega$ meson region is shown on a linear scale. The data are normalized to the simultaneously measured p+p elastic events.}
\label{Lfig1}
\end{figure}

The identified single-electron and single-positron tracks  were combined into opposite-sign pairs. However, many of these pairs represent combinatorial background (CB) which is mostly due to uncorrelated pairs from multi-pion Dalitz decays and correlated ones from $\pi^{0}\to \gamma\gamma$ accompanied by photon conversion in the detector material and/or from Dalitz decays. The combinatorial background was reduced by using an opening angle cut of $\alpha_{ee}>9^{o}$ between the reconstructed lepton tracks and a condition on quality criteria of the track fitting algorithm \cite{Aga:2009}. In addition a momentum cut of 0.08 $<$ p[GeV/c]  $ < $ 2 for each lepton was applied. The combinatorial background was formed from the sum of the reconstructed like-sign invariant-mass distributions, $dN^{++}/dM_{ee}$ and $dN^{--}/dM_{ee}$. The like-sign pairs were subjected to the same selection criteria as the unlike-sign ones. Both, the unlike sign invariant-mass distribution  and the CB, were corrected for the detector and reconstruction inefficiencies on a pair-by-pair basis, defined as the product of single-lepton efficiencies deduced from dedicated Monte-Carlo events embedded into real events \cite{Aga:2009,Aga:2008,Aga:2007}. The geometrical pair acceptance (the acceptance matrix) of the HADES detector was obtained in a similar way from single-lepton acceptances defined as a function of the lepton momentum, polar and azimuthal emission angles \cite{Aga:2009}.

The final signal pair distribution inside the geometrical acceptance of the HADES, shown in Fig.~\ref{Lfig1}, is the result of a subtraction of the CB from the unlike-sign invariant-mass distribution. Both spectra are normalized to $N_{el}^{acc}/\sigma_{el}^{acc}$, where $N_{el}^{acc}$ and $\sigma_{el}^{acc}$ denote the measured yield of the p+p elastic scattering and the differential elastic cross section \cite{Elas:1971} inside the acceptance of HADES, respectively.

The low-mass region of the spectrum in Fig.~\ref{Lfig1} is dominated by Dalitz decays of neutral mesons ($\pi^{0}$, $\eta$, $\omega$), as well as by the Dalitz decay of the $\Delta(1232)$ resonance (see below). The evident peak around the pole mass of the $\omega$ meson corresponds to its direct decay into $e^{+}e^{-}$ pairs. However, this mass range contains also pairs stemming from the direct decays of the $\rho$ meson. In total, 6.1$\times10^{4}$ signal pairs, 5.4$\times10^{4}$ of them in the region below  0.15 GeV/$c^{2}$, were reconstructed. The number of pairs in the mass range between 0.71 GeV/$c^{2}$ and 0.81 GeV/$c^{2}$, which corresponds to the $\pm$3$\sigma$ interval around the reconstructed $\omega$ peak, amounts to 260. In the inset of Fig.~\ref{Lfig1}, the $\omega$ peak is shown on a linear scale. In order to estimate the mass resolution, this peak was fitted with a Gaussian distribution plus a polynomial for the underlying continuum. The obtained mass resolution, $\sigma$/$M^{\omega}_{pole}$, is about 2$\%$.
\vspace{8mm}

\section {Comparison to three models}
\label{sec:Models}
\subsection{PYTHIA+PLUTO, UrQMD, HSD}
\label{subsec:pythia}

For better understanding of the inclusive hadron production in 3.5 GeV p+p interactions, we compare in this section the experimentally measured  distributions to the results from the PYTHIA~\cite{PYTHIA:web}, UrQMD~\cite{UrQMD:1998,UrQMD:web} and HSD~\cite{HSD:web,Brat:2008} event generators.  

PYTHIA, as well as HSD at this energy, use a Monte Carlo realization of the Lund string fragmentation model, where the assumption of a linear confinement potential between the quark and antiquark is taken as a starting point. Although the latter is usually used to describe the multi-particle production in the high-energy regime, it was successfully applied, with some additional adjustments, called tunes, to reproduce the experimental data at low energies.
In particular, since PYTHIA is not predictive in assigning the spin to the newly created quark and antiquark pairs from two adjacent string break-ups, it has a tunable parameter which can be adjusted in order to get vector meson multiplicities in accordance with our measured vector meson yields. Recently such tunes have been obtained by the Giessen group, in particular for the p+p data at 3.5 GeV \cite{Gall:2009,Janus:2010}. Our tuned values are the same, apart from two parameters \footnote{In our simulation the values PARP(91)= 0.44 and PARJ(21)= 0.36 have been used.}.
 
A large fraction of the hadrons produced by fragmentation are unstable and subsequently decay into final states. We do not let the particles decay directly inside PYTHIA, but rather decay them using the PLUTO~\cite{PLUTO:2007} code. The input information obtained from PYTHIA consists thus of the particle multiplicities and their four-momenta.
  
 The simulated dielectron spectra (referred to as the cocktail) can be expressed as the incoherent sum over various sources of dielectrons, such as Dalitz decays of the pseudoscalar mesons $\pi^{o}$ and $\eta$, Dalitz decay of the vector meson $\omega$, Dalitz decay of the $\Delta(1232)$ as well as direct vector meson decays V$\to$ e$^{+}$e$^{-}$ with V=$\rho$, $\omega$. The plain bremsstrahlung contribution is expected to be small at the present beam energy \cite{Burkhard:2006}. Similar cocktails have also been considered in previous attempts to describe the dielectron production in pp, pd and pA collisions \cite{Mosel:1989,Titov1:1993,Mosel:1994,Titov2:1995,Ernst:1998,Thomere:2007,Brat:2008,UrQMD:2008,Janus:2010}. Schematically, the differential distribution of dielectrons with invariant mass M$_{ee}$ can be expressed as a superposition of the above mentioned decay channels
 
\begin{equation}
\frac{d\sigma}{dM_{ee}}=\sum_{i}\sigma_{i}\frac{d\Gamma_{i}}{\Gamma^{tot}_{i}dM_{ee}}.
\label{eq1}
\end{equation}
This expression means that a parent hadron $i$ is created in the p+p collision with cross section $\sigma_{i}$ and decays subsequently, thus generating the distribution $d\Gamma_{i}/dM_{ee}$. The factor $1/\Gamma_{i}^{tot}$ is the inverse of the total width of hadron $i$; together with the partial width for the dielectron decay channel it encodes the branching ratio. Broad resonances, such as the $\rho$ and $\Delta(1232)$, are actually generated at masses $m_{\rho}$ and $m_{\Delta}$ (see (2) below), and the decay distribution depends correspondingly on $m_{\rho, \Delta}$ and M$_{ee}$ (cf. equations (20, 21) in \cite{Brat:2008}). Note that non-strange baryon resonances besides the $\Delta(1232)$ are not included in PYTHIA.  
Moreover, in case of Dalitz decays, the mass dependence of the electromagnetic transition form factors should be considered. The mass dependences of the  electromagnetic transition form factors for the $\pi^{o}$, $\eta$ and $\omega$ mesons are parametrized in PLUTO~\cite{PLUTO:2007} according to \cite{Land:1985,Ernst:1998} in agreement with recent measurements \cite{NA60:2009,Metag:2011}. 
The direct decays of vector mesons are treated within the VMD model \cite{Ko:1996}, while the formulas for the pseudoscalar (P = $\pi^{o}$ or $\eta$) and vector meson Dalitz decays are adopted from \cite{Land:1985,Kriv:2001}.
The Dalitz decay of the $\Delta(1232)$ resonance is simulated using the expression for its differential decay rate  from \cite{Kriv:2001}.
In these calculations, the $N-\Delta$ transition vertex is described by electric, magnetic  and Coulomb form factors, corresponding to three independent helicity amplitudes \cite{Scadron:1972}.  As mentioned in the introduction, in the time-like region the $q^{2}$ dependence of the transition form factors is not measured yet. 
Therefore, we make an approximation by  fixing the form factors at the photon point (real photons) using the measured  radiative decay width of the $\Delta(1232)$ ($\Gamma_{\Delta\to N\gamma}$=0.61-0.7 MeV) \cite{PLUTO:2010,Ernst:1998}. We further neglect the terms with the electric form factor,  as the electric transition is much weaker than the magnetic one \cite{Gourdin:1962}. 
 
It should be further noted that in PYTHIA the $\rho$ and $\Delta(1232)$ resonances are implemented with constant total widths around the resonance pole mass. This treatment is not precise enough outside the resonance pole. Therefore, following the prescription of \cite{Brat:Priv,HSD:web}, we  generate masses of  $\Delta(1232)$, $\rho$ and $\omega$ states inside PYTHIA  according to the relativistic Breit-Wigner distribution
\begin{equation}
\label{BRW}
A(M)=N\frac{2}{\pi}\frac{M^{2}\Gamma_{tot}}{\left ( M^{2} - M_{R}^{2} \right )^{2} + (M\Gamma_{tot}(M))^{2}},
\end{equation} 
with mass dependent total width $\Gamma_{tot}(M)$ in case of $\Delta(1232)$ and $\rho$, and a constant total width at the pole for the narrow $\omega$ meson state. This mass dependence of the total width in case of the $\Delta(1232)$ baryon resonance is calculated from its dominant decay channel into pion and nucleon final states with the cutoff parametrization of \cite{Manley:1991} 
\begin{equation}
\label{Manley}
\Gamma_{tot}^{\Delta}(M_{\Delta}) \simeq \Gamma_{\Delta \to \pi N} = \Gamma_{pole}\frac{M_{pole}}{M_{\Delta}}\left(\frac{q}{q_{pole}}\right)^{3}\frac{\delta^{2}+q_{pole}^{2}}{\delta^{2}+q^{2}},
\end{equation} 
where $\delta$ = 0.197 GeV,  M$_{\Delta}$ is the actual mass of the $\Delta(1232)$, M$_{pole}$ is its pole mass and $\Gamma_{pole}$ is its pole width \footnote{In this paper c=1 units are used for the formulas.}.  Furthermore, q and q$_{pole}$ denote the pion 3-momenta in the rest frame of the $\Delta(1232)$ with mass M$_{\Delta}$ and M$_{pole}$, respectively.
In \cite{Kemp:2009} the authors investigate the effect of different cutoff prescriptions \cite{Manley:1991,Monitz:1984}, in particular on the dielectron spectra for high masses. The resulting uncertainties are larger than a factor of 3.  

The mass dependence of the $\rho$ total width is parametrized according to \cite{Brat:2008}.

The normalization constant in Eq.~(\ref{BRW}), $N$,  is chosen such that $\int_{min}^{max}{A(M)dM}=1$, where $max$ is fixed at 2 GeV and $min$ is taken to be 2m$_{\pi}$, 3m$_{\pi}$ and m$_{\pi}$+m$_{N}$ for $\rho$, $\omega$ and $\Delta(1232)$ correspondingly.

In contrast to PYTHIA and HSD, the transport model UrQMD \cite{UrQMD:2008} uses a resonance $($R$)$ excitation mechanism for the production of particles via two-nucleon $($N$)$ reactions of the type NN $\to$ NR, NN $\to$ RR. Resonances with masses up to 2.2 GeV for the N* and 1.95 GeV for the $\Delta$ are included \cite{UrQMD:1998}. The production matrix elements for resonances are obtained from the experimental data on $\pi$, $\eta$ and $\rho$ production, when available. 
The probabilities for the resonances to decay  into specific channels are then given by the corresponding known branching ratios. Even though in UrQMD the excitation of many baryon resonances are used, only the Dalitz decay of the $\Delta(1232)$ isobar is explicitly included. The $e^{+}e^{-}$ contribution from decays of higher-lying resonances is included via their $\rho$ decay branches (see table 3.4 in \cite{UrQMD:1998}). However, this approach leads to an overestimation of the e$^{+}$e$^{-}$ production from the $\rho$ decays. On the other hand, estimates of the e$^{+}$e$^{-}$ yield  from the Dalitz decays of higher resonances indicate smaller contribution as compared to the one from  $\Delta(1232)$ and $\eta$ decays for M$_{ee} < 0.55$ GeV/c$^{2}$ \cite{Mosel:Res,Wolf:2003,Wolf:2005}.  Dalitz decays of higher-lying baryon resonances have also been investigated in \cite{eVMD:2002}. Checking the validity of this approach for describing the experimental data is a subject of ongoing HADES activities. 

In general, the reaction pp $\to$ e$^{+}$ e$^{-}$ X is fully described by three independent degrees of freedom (neglecting the internal degrees of freedom, like helicity angles of virtual photons) which can be selected in a variety of ways. It is important that a given event generator describes the experimentally measured distributions in all degrees of freedom.
Therefore we present here the comparison of pair invariant mass, transverse momentum and rapidity distributions  to the corresponding distributions from simulated PYTHIA, UrQMD and HSD events as discussed above. In doing so, the acceptance matrices mentioned above are used, i.e. the comparison is performed inside the HADES acceptance.
Furthermore, momenta of leptons are smeared in the simulation in order to take into account our finite detector resolution. The smearing functions are obtained by propagating simulated e$^{+}$ and e$^{-}$ tracks through the detector setup using the Geant package \cite{HGeant:2004,Geant:2004}, hence taking into account the interaction of leptons with the detector material as well. This is in particular visible in the simulated $\omega$ peak shape, where the tail towards low masses is due to the energy loss of electrons via electromagnetic radiation (bremsstrahlung). On the other hand, the ionization (collisional) energy loss of electrons shifts the pole position of the reconstructed omega peak by 1 $\%$ downwards.

 \subsection{Invariant mass distribution}
 \label{subs:invmass}
  
  \begin{figure}
    \centering
    \subfigure[]
      {\includegraphics[width=0.8\linewidth,clip=true]
      {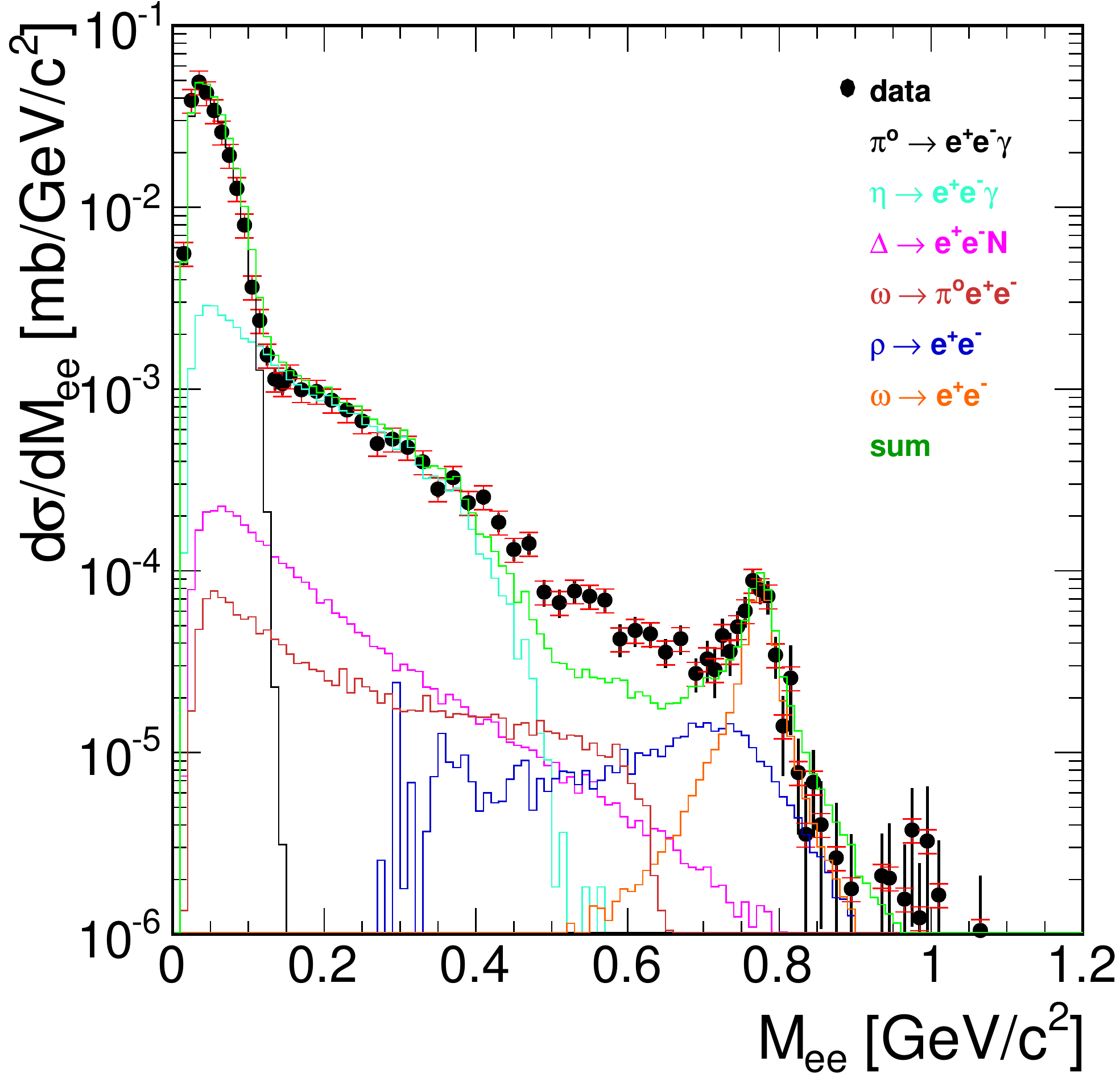}
       \label{invmassP}
      }
      
      \subfigure[]
      {\includegraphics[width=0.8\linewidth,clip=true]
      {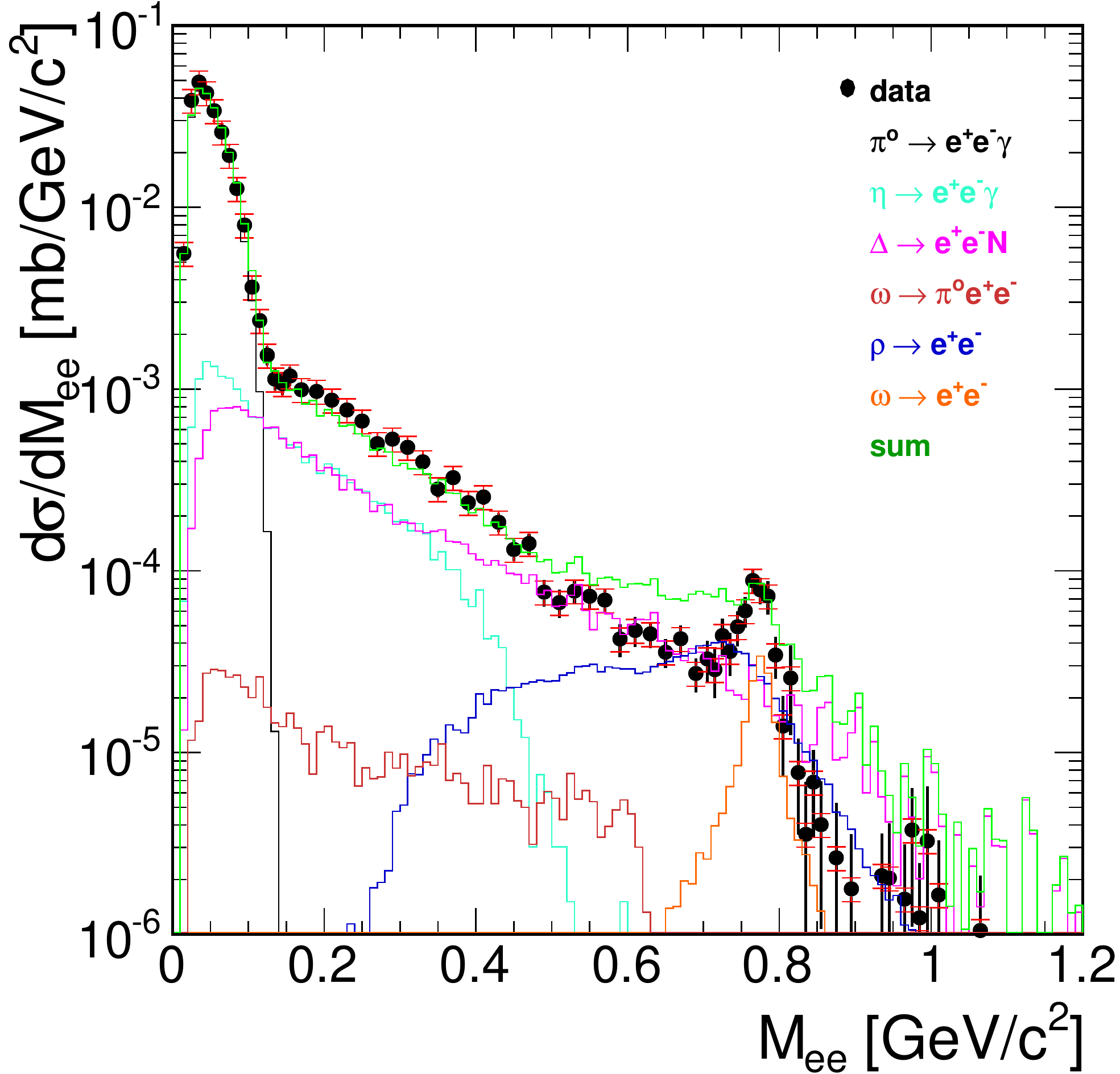}
       \label{invmassU}
      }
            
       \subfigure[]
      {\includegraphics[width=0.8\linewidth,clip=true]
      {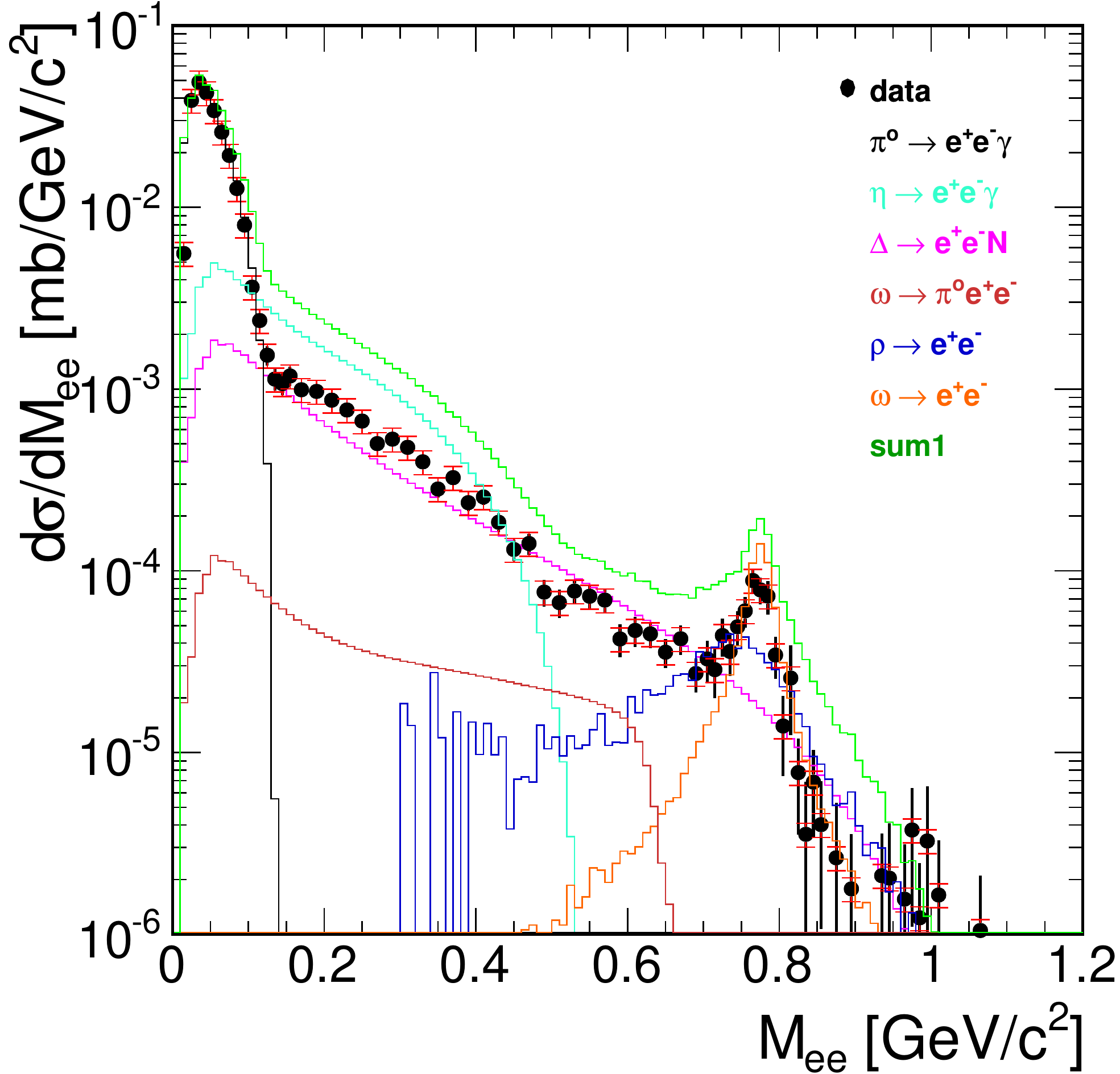}
       \label{invmassH}
      }
     
    \caption{(color online) The HADES data for the p(3.5 GeV) + p $\to$ e$^{+}$ e$^{-}$ X reaction, compared to a simulated cocktail from the a) PYTHIA, b) UrQMD and c) HSD event generators.}
    \label{figInvMass}
\end{figure}
 
The comparison of the experimentally measured invariant mass distribution of $e^{+}e^{-}$ pairs to the PYTHIA+PLUTO  results is presented in Fig.~\ref{invmassP}.  
The simulated cocktail (the sum of the different cocktail contributions is plotted as a green curve) reproduces results of GiBUU \cite{Janus:2010} and describes the data reasonably well except for the mass range around 0.55 GeV/c$^2$ where the yield is underestimated. The latter deviation is not too surprising, as the virtuality of the photon $\gamma^{*}\to e^{+}e^{-}$ reaches quite high values, and therefore it is not guaranteed that the $\Delta(1232)$ form factors fixed at the photon point are still valid.  Furthermore, contributions of higher $\Delta$ and $N^{*}$ resonances are not included which might also lead to some deficit. 
   
In Fig.~\ref{invmassU}, we compare our results to the UrQMD \cite{UrQMD:1998} predictions. In this case, the simulated cocktail clearly underestimates the contribution from $\eta$ and $\omega$ mesons, while the contribution from the $\Delta(1232)$ Dalitz and $\rho$ decays are too strong. The latter might be due to the large R$\to$N$\rho \to$ Ne$^{+}$e$^{-}$ couplings as mentioned already above. The small $\eta$ contribution can be related to the fact that in UrQMD $\eta$ meson production is mediated by the N*(1535) excitation only, which is known to be important for the exclusive channels.  For the inclusive production, however, channels with one and two additional pions are probably important as well. 

Fig.~\ref{invmassH} illustrates the comparison of the same experimental data to the HSD results. In this case, the simulated cocktail has too strong contributions for all components, except the $\pi^{o}$. The comparison of Figs.~\ref{invmassP}, ~\ref{invmassU} and ~\ref{invmassH}  points to a lack of understanding of the relative strengths of the $\Delta(1232)$ and low-mass $\rho$ contributions. For instance, as discussed in \cite{Aga:2010}, implementation of the Iachello model \cite{Iacello:2005} enhances dielectron yield in the high mass region. This has also been demonstrated in \cite{gsi_annual:2010,Janus:2010}.  
We expect hence that our data can help to clarify the issue of form factors in the $\Delta(1232)$ Dalitz decay and the role of other baryon resonances, as mentioned in the introduction.

    \begin{figure}
    \centering
    \subfigure[]
      {\includegraphics[width=1.02\linewidth,clip=true]
      {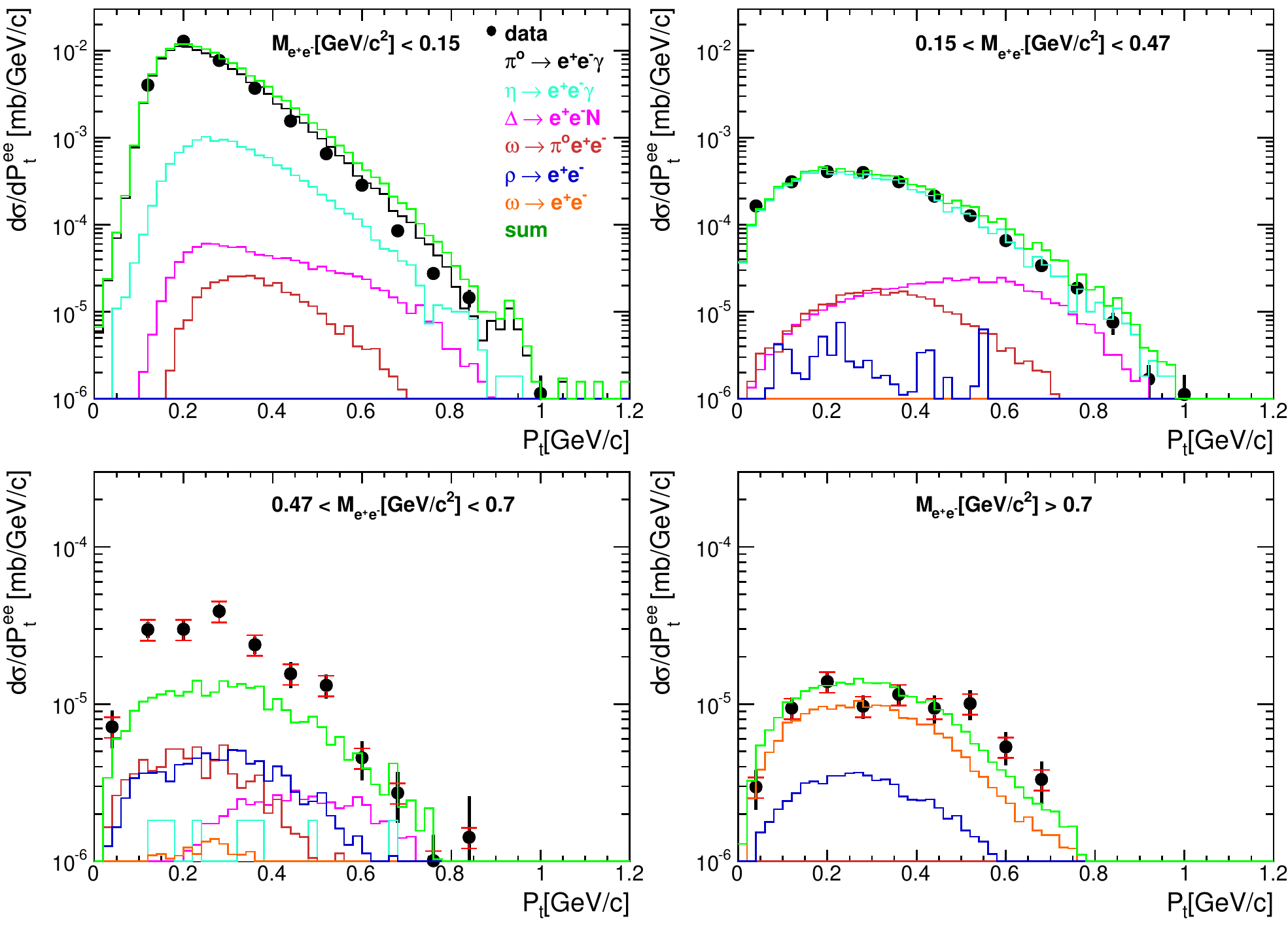}
       \label{figPtP}
      }
      \subfigure[]
      {\includegraphics[width=1.02\linewidth,clip=true]
      {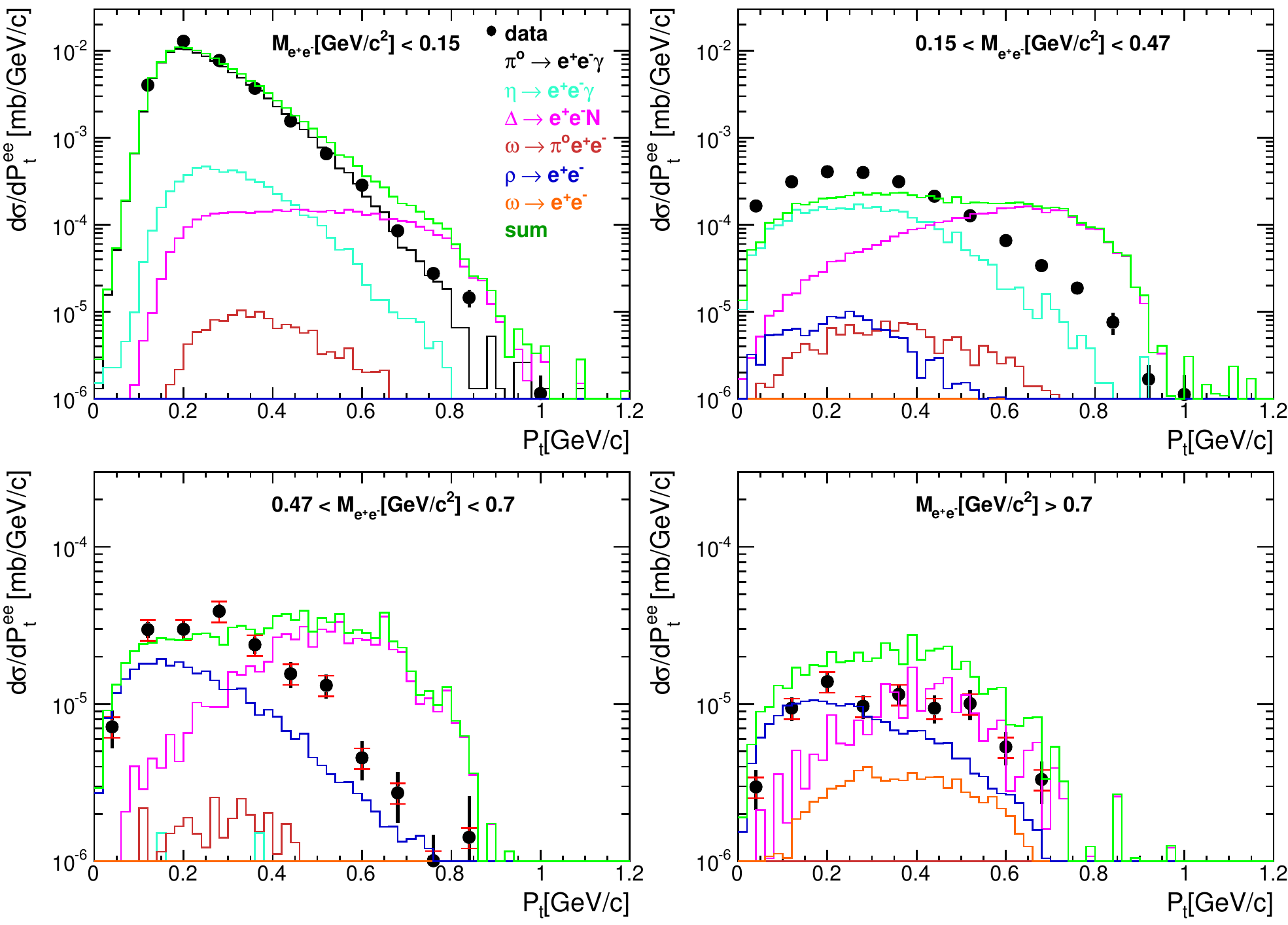}
       \label{figPtU}
      }      
       \subfigure[]
      {\includegraphics[width=1.02\linewidth,clip=true]
      {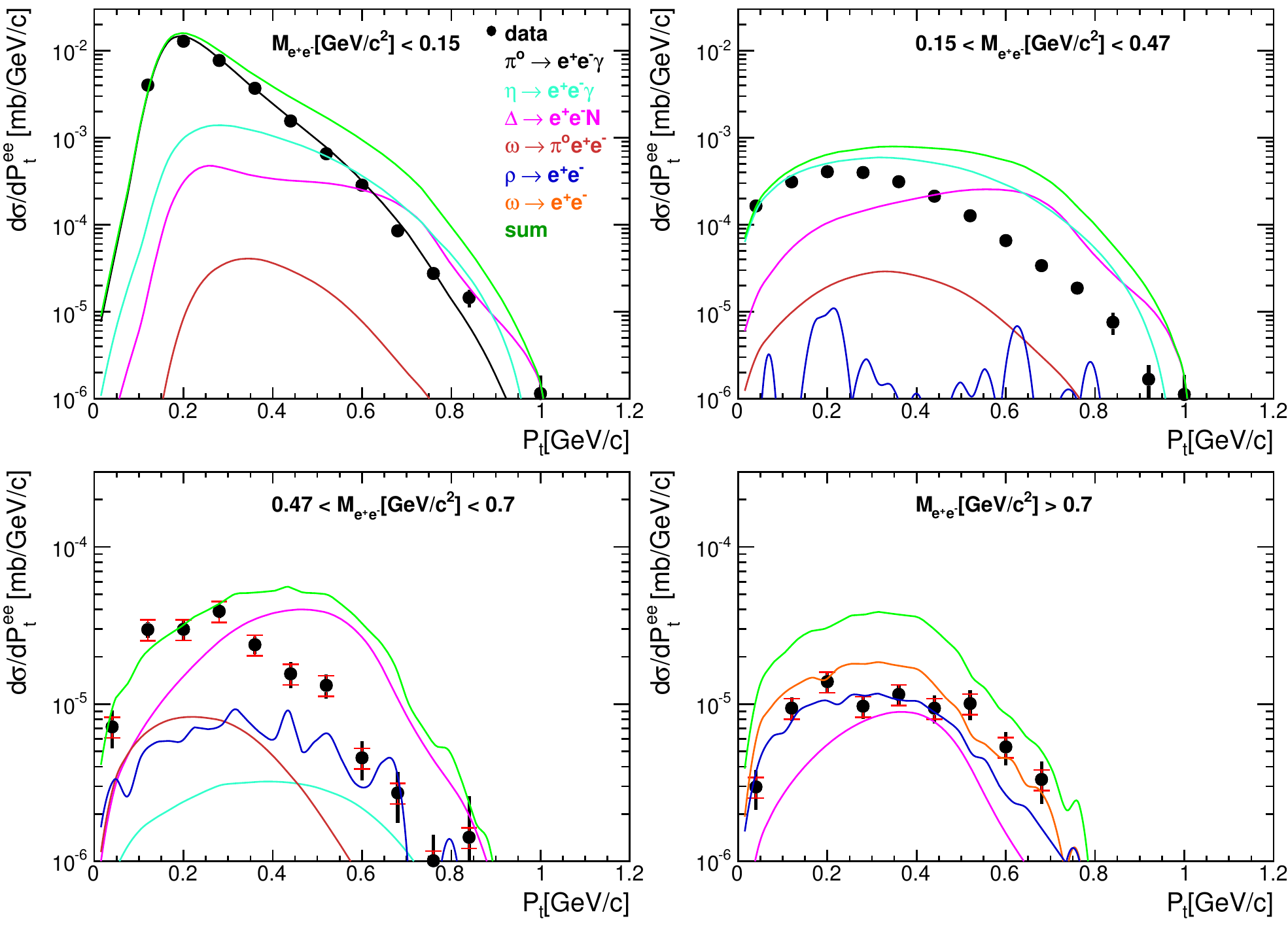}
       \label{figPtH}
      }
    \caption{(color online) Comparison of experimental $p_{t}$ distributions to the a) PYTHIA, b) UrQMD and c) HSD events for different  $e^{+}e^{-}$ invariant mass ranges as indicated.}
    \label{figPt}
\end{figure}

 \subsection{Transverse momentum distributions}
 \label{subs:pt}
 
The $e^{+}$e$^{-}$ pair transverse momentum $p_{t}$ distributions  for different invariant mass bins inside the acceptance of HADES are presented in Figs.~\ref{figPtP}, ~\ref{figPtU}, ~\ref{figPtH} and compared  with the respective results from PYTHIA+PLUTO, UrQMD and HSD calculations.  In the mass range M$_{e^{+}e^{-}}$[GeV/c$^{2}$] $   < $ 0.15, dominated by the contribution from dielectrons stemming from the $\pi^{o}$ Dalitz decay, the experimental $p_{t}$ distributions are in reasonable agreement with the simulated $p_{t}$ distributions from PYTHIA both in shape and absolute yield. The next  mass range of  0.15 $ <  $ M$_{e^{+}e^{-}}$[GeV/c$^{2}$] $ < $ 0.47  constrains the cross sections of the $\eta$ meson and the $\Delta(1232)$ isobar. In this mass interval, the low-$p_{t}$ part of the spectra is populated mainly by the pairs originating from $\eta$ Dalitz decays. The high-$p_{t}$ part contains in addition a substantial contribution from the $\Delta(1232)$ Dalitz decay. Again, tuned PYTHIA reproduces the experimental data, while UrQMD is low by a factor of 2 at low $p_{t}$ and a factor of 5 too high at large $p_{t}$. The HSD results in this mass bin overestimate the experimental results by large factors at high $p_{t}$. In the mass interval of 0.47 $<$ M$_{e^{+}e^{-}}$[GeV/$c^{2}$]  $< $ 0.7 the experimental data cannot be satisfactorily described by any of these models. The $p_{t}$ distribution in the bin M$_{e^{+}e^{-}}$[GeV/$c^{2}$]  $ > $ 0.7, dominated by direct decays of the vector mesons, is again overestimated by the UrQMD and HSD event generators.

 \subsection{Rapidity distributions}
\begin{figure} 
    \centering
    \subfigure[]
      {\includegraphics[width=1.02\linewidth,clip=true]
      {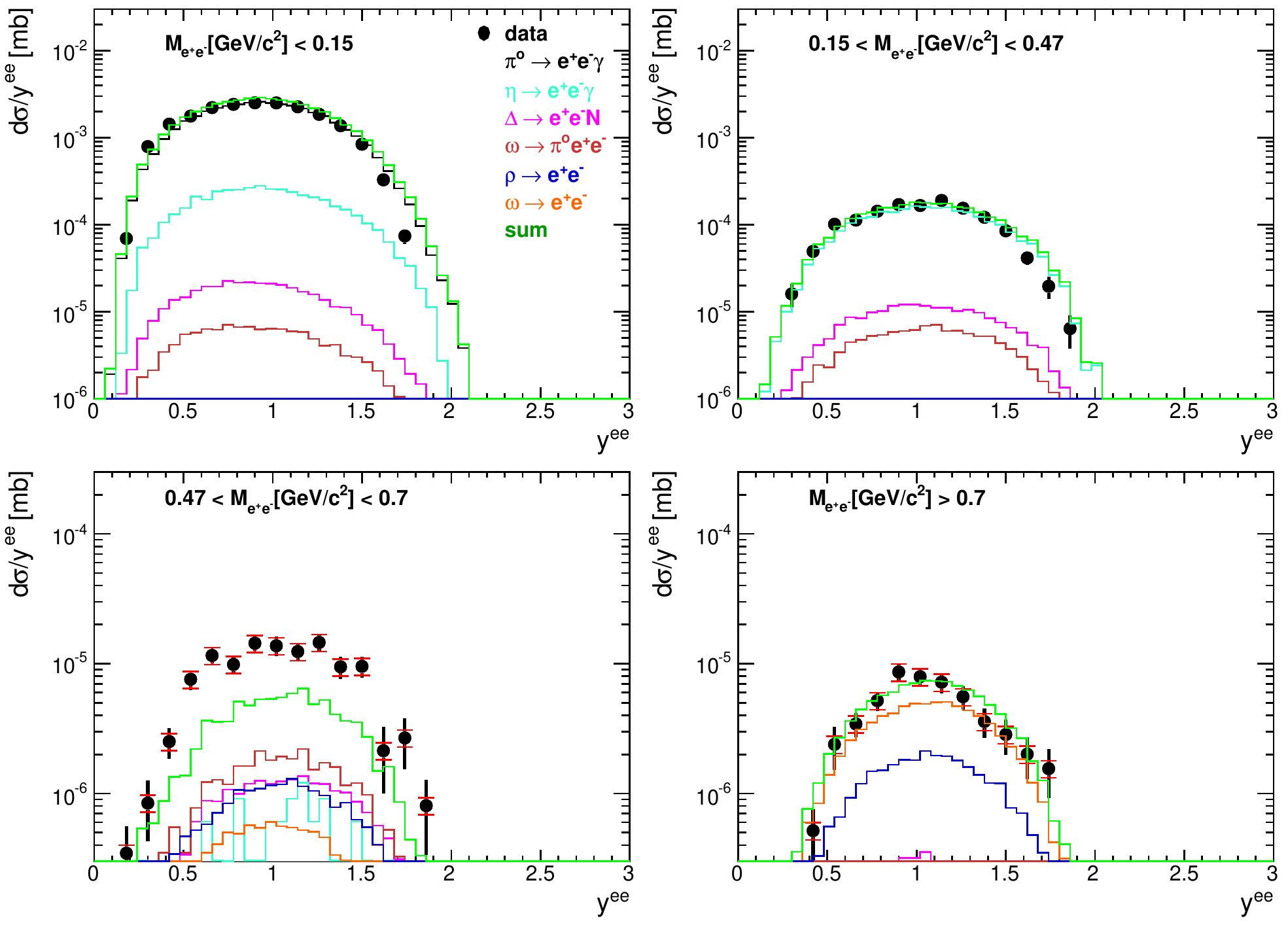}
       \label{figRapP}
      }
      \subfigure[]
      {\includegraphics[width=1.02\linewidth,clip=true]
      {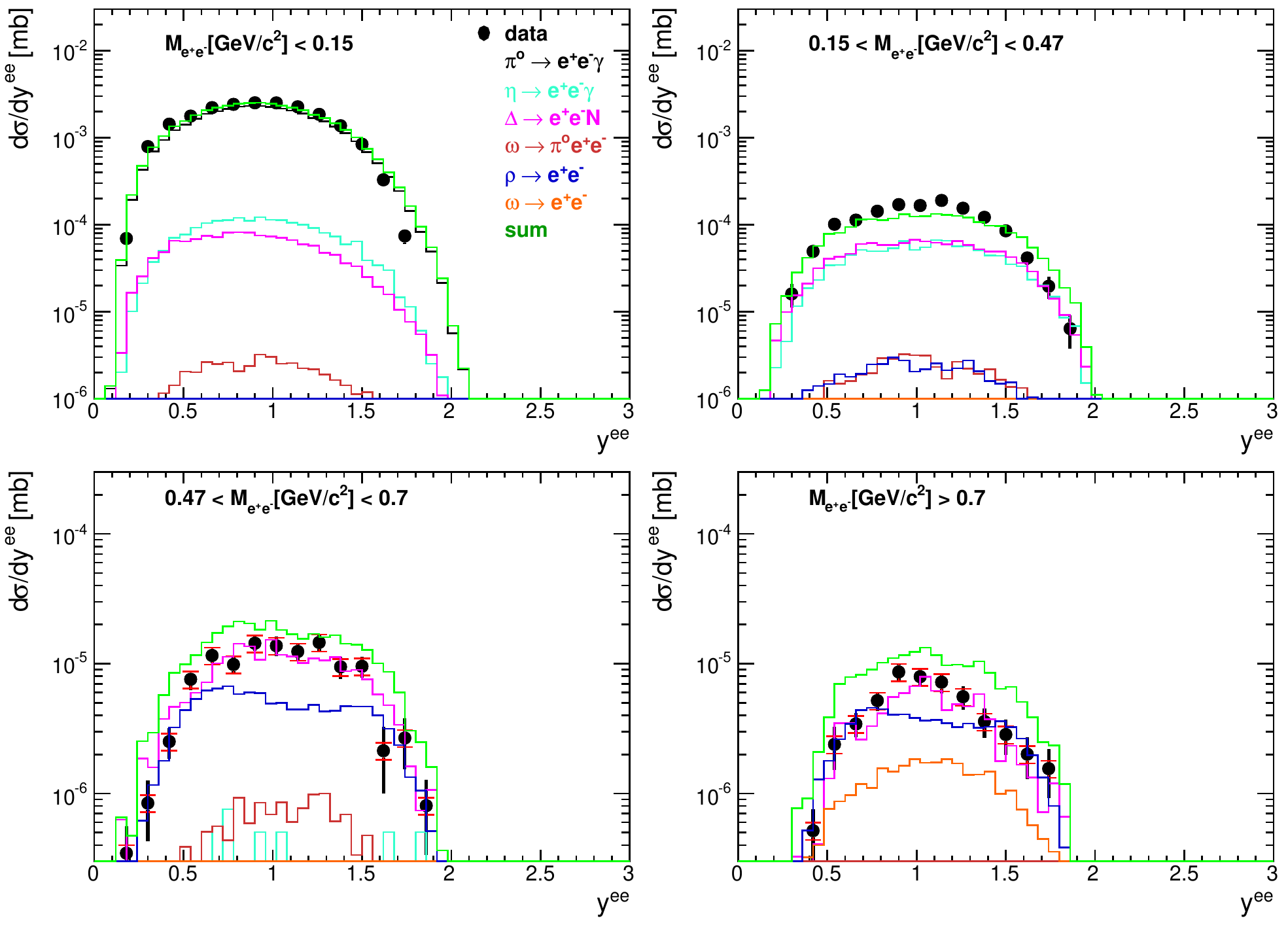}
       \label{figRapU}
      }
        \subfigure[]
      {\includegraphics[width=1.02\linewidth,clip=true]
      {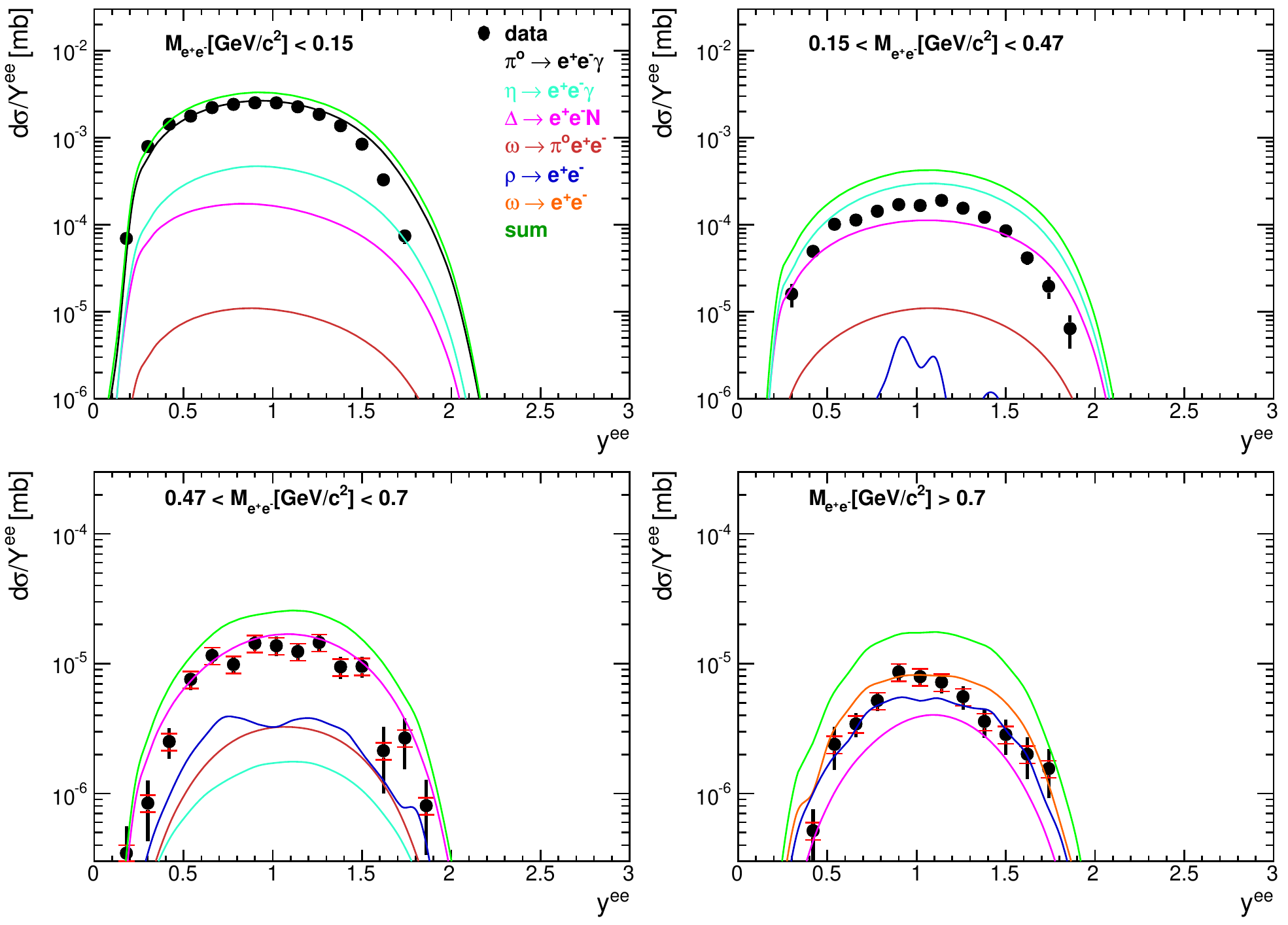}
       \label{figRapH}
      }
    \caption{(color online) Comparison of experimental rapidity distributions to the a) PYTHIA, b) UrQMD and c) HSD events for different $e^{+}e^{-}$ invariant mass ranges as indicated.
   }
    \label{figRap}
\end{figure}

 In contrast to the $p_{t}$ distributions, the rapidity distributions do not exhibit  such large differences between the different models (see Fig.~\ref{figRap}). This may originate from the fact that the impact of the large-p$_{t}$ region to these spectra is reduced due to the $p_{t}$ integration in each rapidity bin.

\section{Extraction of the $\pi^{o}$, $\eta$, $\Delta$, $\rho$, and $\omega$ cross sections}
\label{sec:Cross1}

\begin{figure*}
\begin{center}
 {\includegraphics[width=0.8\linewidth,clip=true]
      {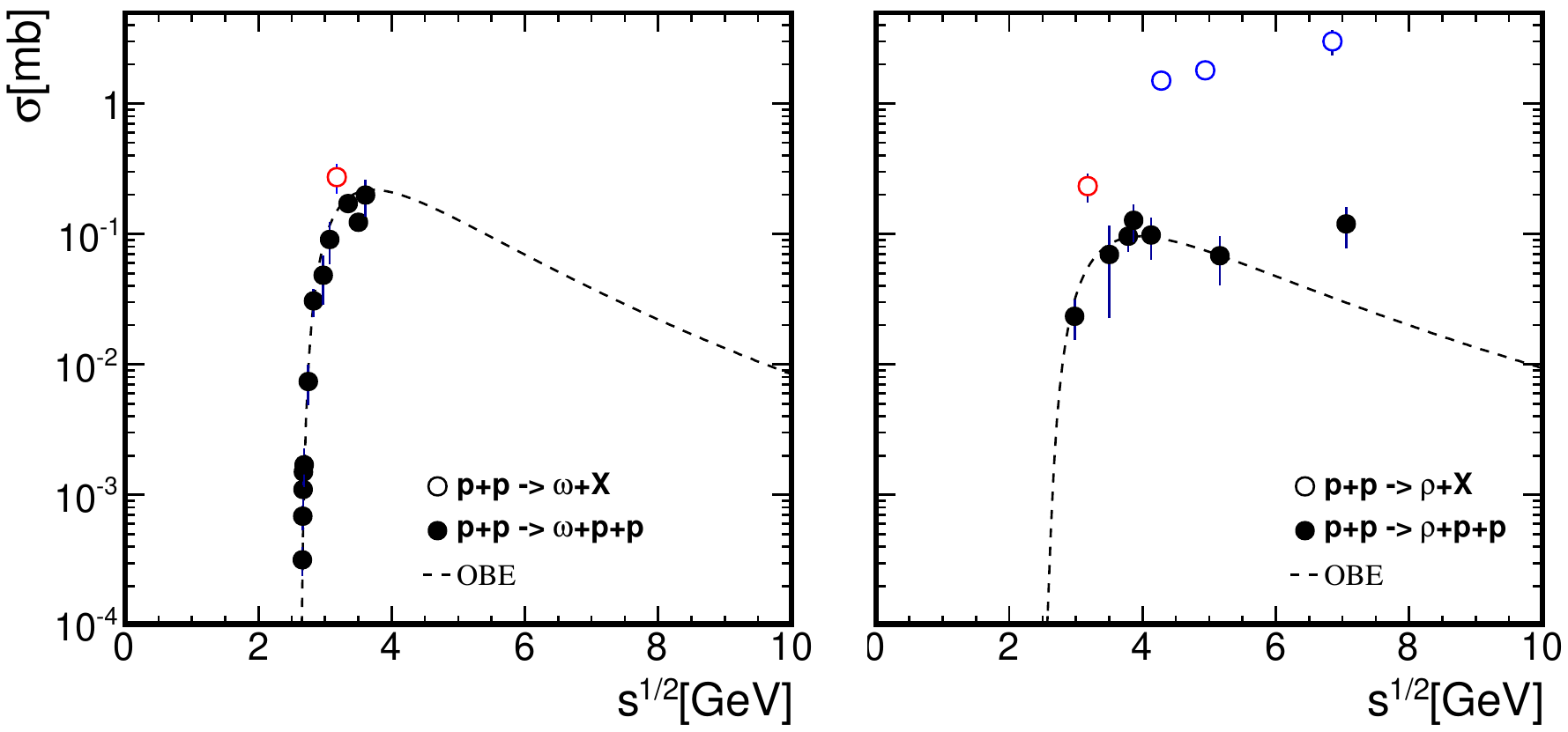}
}

\caption{(color online)  Cross sections for the vector mesons $\omega$ (left panel) and $\rho$ (right panel), in p+p collisions as a function of Mandelstam variable s. Open circles represent inclusive production cross sections, while the full circles correspond to exclusive productions \cite{bornstein:1988}. The dashed curve refers to the OBE calculations for the exclusive channels \cite{OBE:1997}. Cross section values obtained in this work are depicted in red.}
\label{crossVector}
\end{center}
\end{figure*}

\begin{figure*}
\begin{center}
 {\includegraphics[width=0.4\linewidth,clip=true]
      {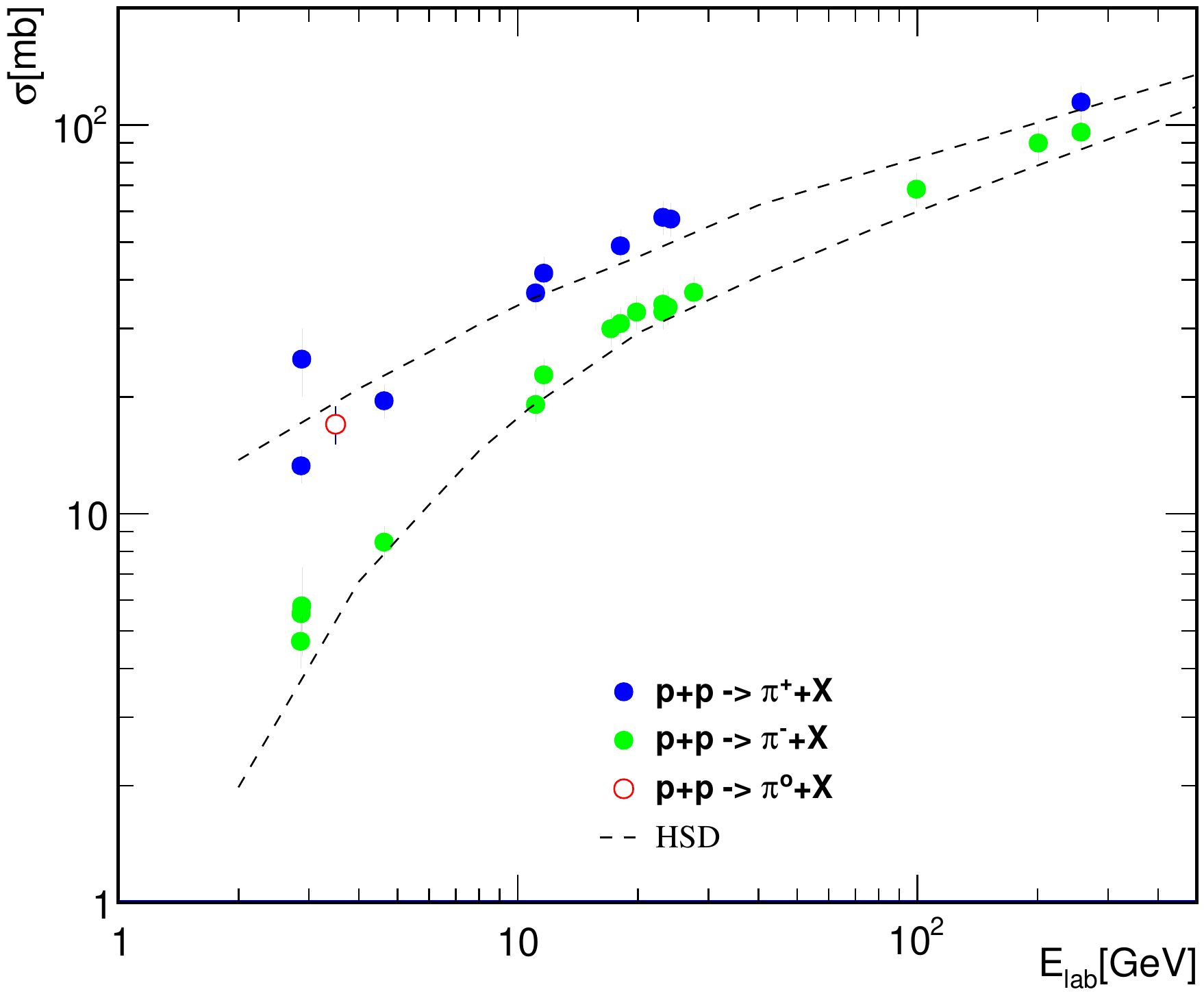}
      }
  {\includegraphics[width=0.4\linewidth,clip=true]
      {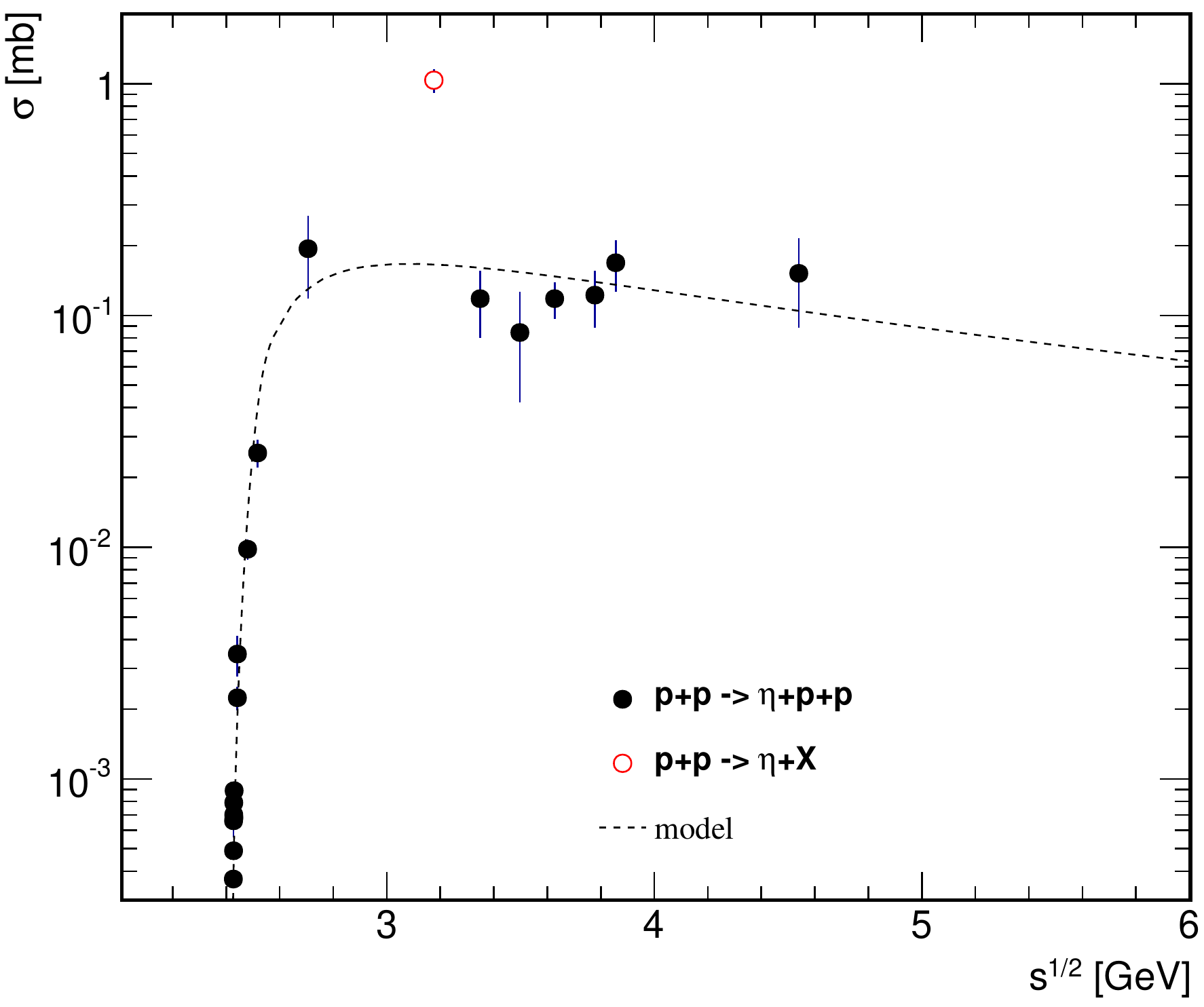}
      }     
\caption{(color online) Cross sections of pions (left panel) and $\eta$ mesons (right panel) in p+p interactions \cite{bornstein:1988,pion1:1973,pion2:1999,eta1:1994,eta2:1993,eta3:1996}. The dashed lines in case of pions refer to the parametrizations used in HSD transport model, while for the $\eta$ meson the line corresponds to its exclusive production through the N*(1535) resonance \cite{Teis:1997}. The full circles in this case illustrate the measured exclusive production cross sections of the $\eta$ mesons in p+p collisions.  Inclusive production cross section values  obtained in this work are depicted with red open circles. 
}
\label{crossPions}
\end{center}
\end{figure*}

From the spectrum presented in Fig.~\ref{invmassP}, the inclusive cross section of the neutral pion production inside the HADES acceptance was estimated from its Dalitz decay channel, $\pi^{0}\to e^{+}e^{-}\gamma$, by integrating the yield in the mass interval between 0 and 0.15 GeV/$c^2$, and by correcting for the branching ratio (1.198$\pm$0.032)$\%$ \cite{PDG:2010}. This result is further extrapolated to full phase space using simulated events. To investigate the model dependence, the extrapolation procedure is repeated for the PYTHIA and UrQMD event generators. The  obtained inclusive neutral pion production cross section in full phase space amounts to 16 $\pm$ 2.6 mb and 18 $\pm$ 2.7 mb, depending on the model used for the extrapolation. The errors for this cross section consist of a statistical part (negligible in the pion region) and a systematic part stemming mainly from the normalization to the elastic p+p cross section and the efficiency correction procedure. The cross section for the $\eta$ mesons were obtained in a similar way by integrating directly its contribution to the simulated cocktail presented in Fig.~\ref{invmassP}, correcting for its Dalitz decay branching ratio (7$\pm$0.7)$\times$10$^{-3}$ \cite{PDG:2010} and extrapolating to full phase space using again UrQMD and PYTHIA+PLUTO simulations. Equivalently, one could directly integrate the experimental data after subtracting the contributions from other sources in the corresponding mass range. In any of these cases there is a contribution from the Dalitz decays of the $\Delta(1232)$ isobar which is also important in the mass range relevant for the $\eta$ meson Dalitz decays \footnote{Note that both, $\Delta^{+}$ and $\Delta^{0}$ states, are taken into account}. This is best seen in the transverse momentum distributions presented in section~\ref{subs:pt}. Indeed, as it is seen from Fig.~\ref{figPtP} the $p_{t}$ distribution corresponding to the invariant mass range of 0.15 - 0.47 GeV/c$^{2}$ essentially fixes the relative contribution of the $\eta$ meson and $\Delta(1232)$ baryon (see section~\ref{subs:pt} for details).  The contribution from the $\omega$ Dalitz decay in this mass range is fixed by its known branching ratio and cross section defined by its direct decay channel. After getting a satisfactory description of the data with the PYTHIA event generator (see Fig.~\ref{invmassP}), the production cross sections for the vector mesons were obtained from their multiplicities in full phase space generated by PYTHIA. 
The acceptance of the spectrometer in the vector meson mass range is high in all dimensions of the phase space. Therefore, to a large extent, the cross sections for vector mesons obtained from PYTHIA are model independent. Our final numbers for the  cross sections  are listed in Table 1. The uncertainties for the  cross sections are calculated based on experimental errors (statistical + systematic).
Fig.~\ref{crossVector} shows a compilation of measured production cross sections of vector mesons ($\omega$, $\rho$) in p+p reactions at different energies. Similar plots for different pion species and the $\eta$ meson are shown in Fig.~\ref{crossPions}, where we use the average values of the cross sections obtained with PYTHIA and UrQMD extrapolations listed in Table 1(for details see the text above).

\begin{table*}
\begin{center}
\begin{tabular}{|c|c|c|c|c|c|}
\hline\hline  & $\pi^{o}$ & $\eta$ & $\Delta^{o,+}$ & $\rho$ & $\omega$ \\ [0.5ex]
\hline  $\sigma_{i}$ [mb]: & 18$\pm$ 2.7 (16 $\pm$ 2.6) &  1.14$\pm$0.2 (0.93$\pm$ 0.14)& 7.5 $\pm$ 1.3 & 0.233 $\pm$ 0.06 &  0.273 $\pm$ 0.07\\ [0.5ex]
\hline\hline	
\end{tabular}
\end{center}
\label{default}
\caption{Inclusive cross sections obtained for different particles. For $\pi^{o}$ and $\eta$ mesons the uncertainty due to model dependence has been included as explained in the text. The values corresponding to the extrapolation with the UrQMD model are presented in brackets.}
\end{table*}

\vspace{5mm}

\section {Direct decays of the $\eta$ meson}
\label{sec:etaDir}
\begin{figure}
\begin{center}
{\includegraphics[width=0.99\linewidth,clip=true]
{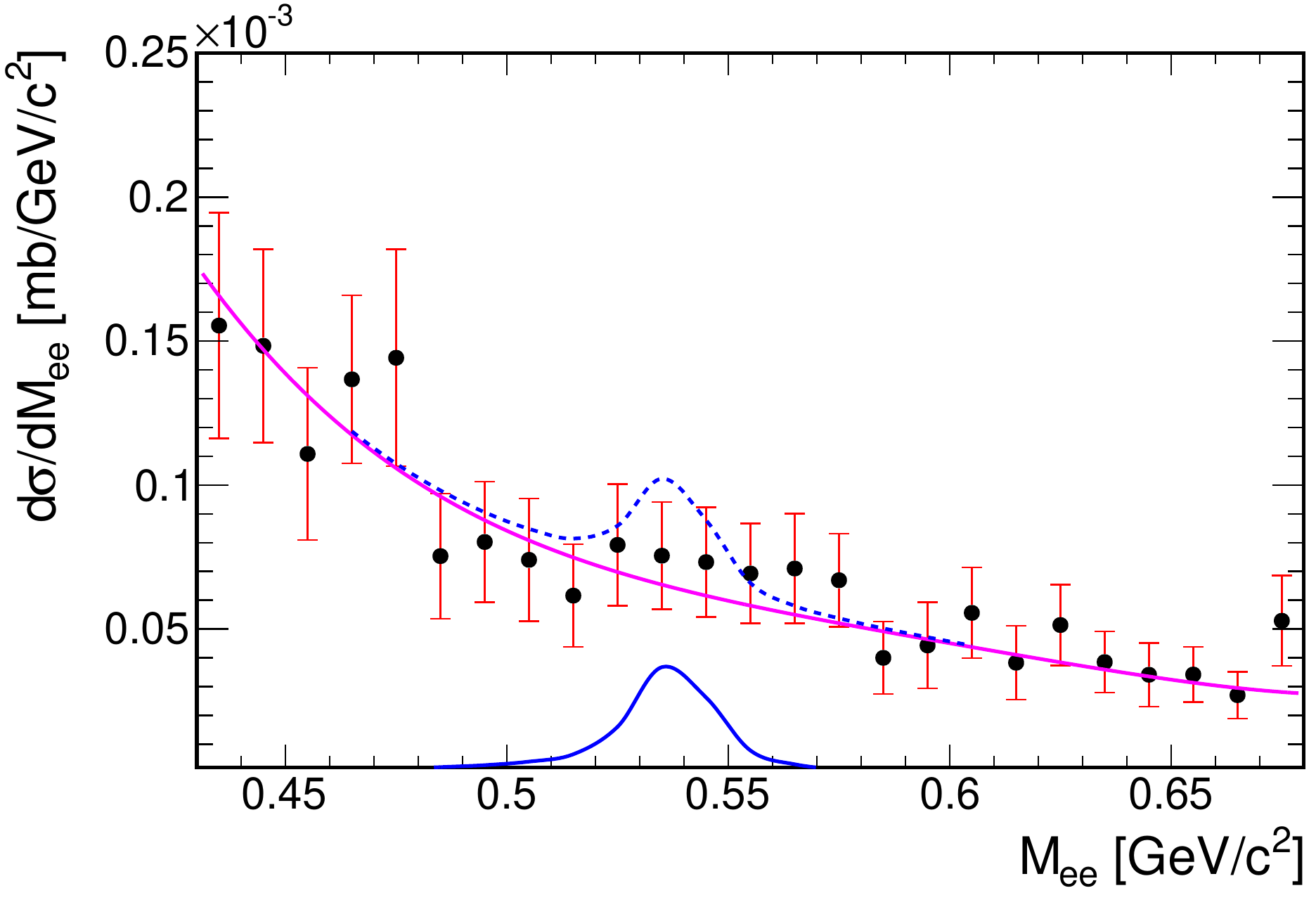}
}
\caption{(color online) Invariant mass distribution of $e^{+}e^{-}$  pairs in the $\eta$ meson mass range. The experimental data is fitted with a polynomial distribution (magenta curve) by excluding the range of $m_{\eta}\pm 3\sigma$ around $\eta$ meson pole mass. The invariant mass distribution of e$^{+}$e$^{-}$ pairs from simulated direct dielectron decays of the $\eta$ meson (blue curve) is presented on top of the background (dashed-blue curve).}
\label{figEtaDir}
\end{center}
\end{figure}

Using the data described above we can derive an upper limit for the branching ratio of the direct $\eta$ meson decay $\eta \to e^{+}e^{-}$.   
Fig.~\ref{figEtaDir} shows the invariant mass distribution of the $e^{+}e^{-}$ pairs in the $\eta$ meson mass range. The experimental data points are fitted with a polynomial background function, excluding some range around the $\eta$ meson mass. There is no visible indication of a peak structure from the direct decays of $\eta$ mesons. Indeed, performing the Kolmogorov-Smirnov test one gets a 90$\%$ probability for the consistency of the background function and the data points. However, one can still estimate an upper limit for this decay process from the data using the method of Feldman and Cousins \cite{Feldman:1998}. In order to test the robustness of the obtained results, five different ranges of the background function have been tested. Moreover, the same procedure was repeated for different polynomial functions. The output of the Feldman Cousins method is an upper limit for the signal counts with a 90$\%$ confidence level. Using this upper limit for the signal counts and the $\eta$ production cross section, reported in the previous section, an upper limit for the branching ratio of  (4.9+0.7-1.2)$\times 10^{-6}$ is obtained. This value is about 6 times lower than the most recent value of  $2.7\times 10^{-5}$ from \cite{EtaDir:Wasa}.
To demonstrate how the hypothetical $\eta$ peak would look like, we show in Fig.~\ref{figEtaDir} the $\eta$ shape from a simulation on top of the experimental data, using the branching ratio of 4.9$\times 10^{-6}$ and the production cross section of 1.14 mb from Table 1.  The significance of the added hypothetical peak amounts to 5.5 $\%$. 

\section{Summary}
In summary, we reported on a dielectron measurement in p+p collisions at 3.5 GeV projectile kinetic energy. For the first time the inclusive production cross sections for neutral pions, $\eta$, $\omega$ and $\rho$ mesons were determined from dielectron experimental data. The experimental distributions were compared to the results from the PYTHIA, UrQMD and HSD event generators, which use different physics assumptions to generate the parent hadrons decaying subsequently into e$^{+}$e$^{-}$ at this projectile energy. With some minor tunes, PYTHIA+PLUTO results describe the experimentally observed distributions in a better way than the resonance production picture used in UrQMD. We hope that our data stimulate further work pinning-down the issue of formfactors in $\Delta(1232)$ Dalitz decays and shed light on the role of other baryon resonances in p+p collisions. 
Using our data, it was demonstrated that the upper bound for the direct $\eta \to e^{+}e^{-}$  decay can be improved by a factor of $\sim$6 compared to the value quoted in \cite{EtaDir:Wasa}.

\section {Acknowledgments}
We would like to thank our theory colleagues, especially Elena Bratkovskaya, Janus Weil, Elvira Santini, Marcus Bleicher and Gyuri Wolf for useful discussions and suggestions.

The collaboration gratefully acknowledges the support by LIP Coimbra,
Coimbra (Portugal): PTDC/FIS/113339/2009, SIP JUC Cracow, Cracow (Poland): N N202 286038 28-JAN-2010 NN202198639 01-OCT-2010, FZ Dresden-Rossendorf (FZD), 
Dresden (Germany): BMBF 06DR9059D, TU M¬unchen, Garching (Germany) MLLM¬unchenDFG EClust: 153VH-NG-330, BMBF 06MT9156 TP5 TP6, GSI TMKrue 1012, NPI AS CR, GSI TMFABI 1012, Rez, Rez (Czech Republic): MSMT LC07050 GAASCR IAA100480803, USC - S. de Compostela, Santiago de Compostela (Spain): CPAN:CSD2007-00042, Helmholtz alliance HA216/EMMI.

 \bibliographystyle{}
 \bibliography{}
%

\end{document}